\documentclass[final,5p,times,twocolumn]{elsarticle} % Removed authoryear option
\usepackage{graphicx}
\usepackage{amssymb}
\usepackage{tikz} % TikZ for vector graphics
\usepackage{lipsum}
\usepackage{amsmath}

\usepackage[font=small, labelfont=bf]{caption}  % Provides caption control (single load)
\usepackage{enumitem} % Add this package for custom numbering
\usepackage{placeins}
\usepackage{adjustbox}     % For adjustbox alignment
\usepackage{subcaption}    % For subfigures (ensure it's only loaded once)
\usepackage{float}         % For figure placement control
\usepackage{afterpage}
\usepackage{balance}
\usepackage{natbib}
\journal{Nuclear Instruments and Methods in Physics Research Section B}
\biboptions{numbers,sort&compress}

\usepackage[colorlinks=true, citecolor=blue, linkcolor=blue, urlcolor=blue]{hyperref} % Only load hyperref once
\begin{document}
\renewcommand{\bibsection}{%
  \section*{References}
  \noindent\rule{\linewidth}{0.4pt}\vspace{1ex}
}

\begin{frontmatter}

\title{Rutherford Backscattering Spectrometry analysis of the formation of superconducting V$_3$Si thin films}

\author[first]{Fshatsion B. Gessesew}
\author[first]{Manjith Bose}
\author[first]{Kumaravelu Ganesan}
\author[second]{Brett C. Johnson}
\author[first]{Jeffrey C. McCallum}
\cortext[First]{Jeffrey McCallum. Email: \href{mailto:jeffreym@unimelb.edu.au}{jeffreym@unimelb.edu.au}}

\affiliation[first]{organization={School of Physics, The University of Melbourne},%Department and Organization
            city={Melbourne},
            postcode={3010}, 
            state={Vic},
            country={Australia}}
\affiliation[second]{organization={School of Science, RMIT University},%Department and Organization
            city={Melbourne},
            postcode={3001}, 
            state={Vic},
            country={Australia}}     
          
\begin{abstract}
Vanadium silicide, V$_3$Si, is a promising superconductor for silicon-based superconducting (SC) devices due to its compatibility with silicon substrates and its potential for integration into existing semiconductor technologies. 
However, to date there have been only a limited number of studies of the formation of SC V$_3$Si thin films and the associated structural and superconducting properties.
This work aims to explore the structural characteristics and SC properties of V$_3$Si films, paving the way for the development of functional SC devices for quantum technology applications.
We have investigated the formation of V$_3$Si films by directly depositing vanadium (V) onto thermally grown SiO$_2$ on Si, followed by high-vacuum annealing to induce the phase transformation into V$_3$Si. Rutherford Backscattering Spectrometry(RBS) was employed throughout the sample growth process to analyze the material composition as a function of depth  using a \(^4\text{He}^+\) ion beam. Analysis of the RBS data confirmed that the V layer fully reacted with the SiO$_2$ substrate to form V$_3$Si at the interface, in addition to a vanadium oxide (VO$_x$) layer forming atop the V$_3$Si film. The thickness of the V$_3$Si layer ranges from 63 to 130 nm, with annealing temperatures between 750°C and 800°C.  A sharp SC transition was observed at T$_c$ = 13 K in the sample annealed at 750°C, with a narrow transition width $(\Delta T_c) \text{ of } 0.6~\text{K}$. Initial reactive ion etching (RIE) studies yielded promising results for local removal of the (VO$_x$) to facilitate electrical contact formation to the SC layer.
 
\end{abstract}


\end{frontmatter}

\section{Introduction}
\label{introduction}
Over the past decade, A15 structure SC materials have gained significant attention due to their relatively high SC transition temperatures, \cite{bib6} and efforts have been made to fabricate Josephson devices from materials such as Nb$_3$Sn, \cite{bib16} and Nb$_3$Ge \cite{bib17}. Vanadium silicide, V$_3$Si, is an example of an A15-structure SC material, first discovered by Hardy and Hulm in 1954, with a critical temperature (T$_c$) of 17.1 K. \cite{bib1} As a silicide, V$_3$Si is compatible with existing silicon technology \cite{bib2, bib3} and possesses the highest T$_c$ among other silicides, making it an attractive material for the development of silicon-based SC devices. Notably, since it has an s-wave pairing symmetry and isotropic energy gap, it is an excellent candidate for use in devices like Josephson junctions and superconducting quantum interference devices (SQUIDs). \cite{bib4,bib5}

V$_3$Si film can be formed using a phase reaction technique, in which V metal is deposited directly onto Si or SiO$_2$/Si substrates using various deposition methods, such as electron beam evaporation \cite{bib8,bib10}, magnetron sputtering \cite{bib11,bib15}, or molecular beam epitaxy \cite{bib3} followed by annealing under optimal conditions. Annealing V/Si films, where V is directly deposited on Si, usually results in the formation of VSi$_2$, which is not a SC material.\cite{bib7, bib8, bib13} However, Zhang et al. \cite{bib9} have recently demonstrated the formation of V$_3$Si by depositing V onto a 20 nm thick top layer of a silicon-on-insulator (SOI) substrate, with the layer thickness of V being chosen to approximately match the amount needed to give the the correct stoichiometry of V$_3$Si in the fully transformed film. Following heat treatment at specific conditions, the V$_3$Si formed leaving a thin residual layer of unreacted V layer at the surface. 

In the case of V/SiO$_2$/Si film formation, where V is deposited on a SiO$_2$/Si substrate, the formation of V$_3$Si phase initiates at a somewhat higher threshold temperature than is the case for direct reaction of V with Si. In both cases, annealing to form the SC phase must be carried out in a high vacuum environment.\cite{bib7, bib8, bib12,bib13,bib14} Compared to V/Si films, the diffusion rate of Si in V/SiO$_2$/Si films is lower because the Si source originates from SiO$_2$, which must first undergo decomposition before diffusion can occur. Thus, during the annealing of V/SiO$_2$/Si films, two key processes are expected to take place: (i) the decomposition of SiO$_2$ and (ii) the subsequent interaction of the dissociated Si and O with V, leading to the formation of V$_3$Si beneath a VO$_x$ surface layer.\cite{bib10,bib13,bib14} 

In this report, we present our preliminary results on the fabrication of SC devices; mainly, the formation of SC V$_3$Si thin films on thermally grown SiO$_2$ on Si, achieved through interfacial reaction and phase transformation. The results include RBS analysis for compositional and structural characterization, optical micrograph analysis of the sample’s surface, and low-temperature four-terminal resistance measurements to investigate the SC behavior of the V$_3$Si thin film. Furthermore, we report the progress on the application of reactive ion etching (RIE) to remove the phase-transformed VO$_x$ layer, to expose the V$_3$Si phase for electrical contacting. 
\section{Thin film fabrication}
\begin{figure*}[ht] % Use figure* for full-width figure in two-column layout
    \centering
    \begin{tikzpicture}
        % Legend
        \node[rectangle, align=left, anchor=north west] (leg) at (-2, -1) {
            \includegraphics[width=10pt]{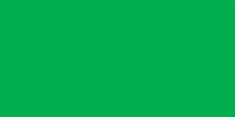} Si \\
            \includegraphics[width=10pt]{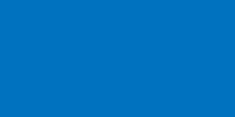} SiO$_2$ \\
            \includegraphics[width=10pt]{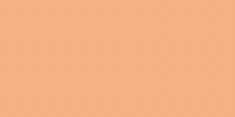} V \\
            \includegraphics[width=10pt]{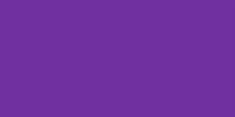} V$_3$Si \\
            \includegraphics[width=10pt]{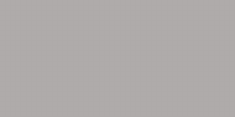} VO$_x$
        };

        % Subfigures
        \node[anchor=north west] (a) at (-0.2,-0.45) {
            \begin{subfigure}[b]{0.2\textwidth}
                \centering
                \includegraphics[width=\textwidth]{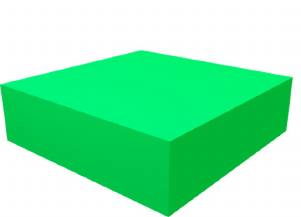}  \vspace{4pt}
                \caption{Si wafer - substrate}
                \label{fig:Si-wafer}
            \end{subfigure}
        };
        
        \node[anchor=north west] (b) at (4.1,0) {
            \begin{subfigure}[b]{0.2\textwidth}
                \centering
                \includegraphics[width=\textwidth]{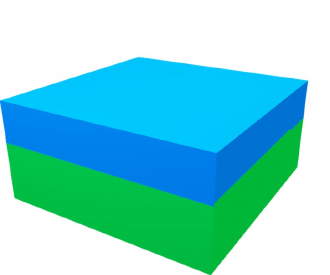}  \vspace{4pt}
                \caption{SiO$_2$ growing}
                \label{fig:SiO2}
            \end{subfigure}
        };
        
        \node[anchor=north west] (c) at (8.17,0) {
            \begin{subfigure}[b]{0.2\textwidth}
                \centering
                \includegraphics[width=\textwidth]{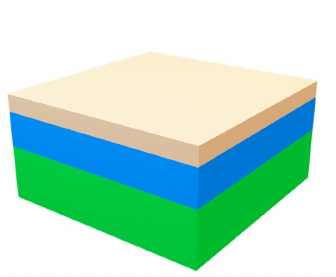}  \vspace{4pt}
                \caption{V deposition}
                \label{fig:V-deposition}
            \end{subfigure}
        };

        \node[anchor=north west] (d) at (12.35,0) {
            \begin{subfigure}[b]{0.2\textwidth}
                \centering
                \includegraphics[width=\textwidth]{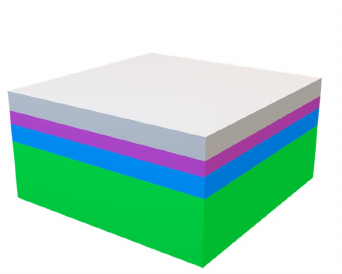}  \vspace{4pt}
                \caption{Anneal}
                \label{fig:annealing}
            \end{subfigure}
        };

        % Subfigure (e) at bottom center
        \node[anchor=north] (e) at (8,-6) {
            \begin{subfigure}[b]{0.8\textwidth}
                \centering
                \includegraphics[width=\textwidth]{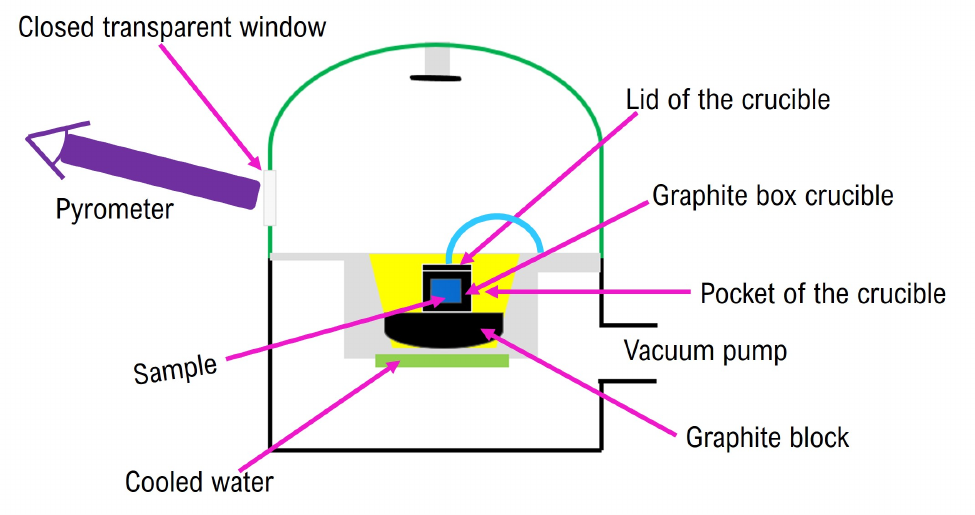} \vspace{5pt}
                \caption{Setup diagram of UHV pyrometer-equipped electron beam annealing system}
                \label{fig:ebeam-annealing}
            \end{subfigure}
        };

        % Arrows
        \draw[->, ultra thick] (3.2,-2.8) -- (4.2,-2.8); % from (a) to (b)
        \draw[->, ultra thick] (7.5,-2.8) -- (8.5,-2.8); % from (b) to (c)
        \draw[->, ultra thick] (11.8,-2.8) -- (12.8,-2.8); % from (c) to (d)
    \end{tikzpicture}
    \caption{V$_3$Si thin film fabrication process. (a) Si wafer (100) substrate. (b) Thermal growth of SiO$_2$. (c) Vanadium deposition. (d) Annealing. (e) Configuration of the temperature controller pyrometer and the UHV electron beam annealer}
    \label{fig:Figure_1}
\end{figure*}

In this study, we synthesized V$_3$Si through the phase transformation of V/SiO$_2$/Si films.
V/SiO$_2$/Si films were prepared as follows: 
First, a 230 nm thick SiO$_2$ layer was grown by thermally oxidizing a Si wafer in a dry oxygen ambient at 1000°C for 5 hours and 15 minutes. The Si wafer was thoroughly cleaned using Piranha/RCA2/HF to remove any contaminants and native SiO$_2$ prior to oxidation. Finally, a 300 nm thick V layer was deposited onto the 230 nm thick SiO$_2$/Si using an electron beam evaporation system with a base pressure of about $1 \times 10^{-7}$ Torr.
V is very reactive at elevated temperatures\cite{bib30} and usually requires annealing to be performed under ultrahigh vacuum (UHV) conditions. Therefore, we employed an in-house developed annealing technique utilizing high vacuum electron beam annealer (HVEBA) equipped with a disappearing-filament pyrometer, and a graphite box crucible, as shown in Figure \ref{fig:Figure_1}(\textcolor{blue}{e}). The Sample was enclosed in a graphite box, within the EBA chamber, which was heated by the electron beam. The temperature of the sample was monitored with a pyrometer from outside of the chamber. 
Figure \ref{fig:Figure_1}(\textcolor{blue}{a-d}) shows the process employed to synthesize V$_3$Si films. The process begins with preparing a Si substrate from a cleaned Si wafer (Figure \ref{fig:Figure_1}\textcolor{blue}{a}), followed by the thermal growth of a SiO$_2$ layer (Figure \ref{fig:Figure_1}\textcolor{blue}{b}). V is then deposited onto the SiO$_2$/Si substrate (Figure \ref{fig:Figure_1}\textcolor{blue}{c}), and finally, the sample is annealed to induce the formation of V$_3$Si (Figure \ref{fig:Figure_1}\textcolor{blue}{d}).
 Thin film samples were annealed at 750 °C  or 800 °C for 30 minutes to form a layered structure consisting of a residual layer of unreacted SiO$_2$ with a thickness that depends on the duration of the anneal, the V$_3$Si layer formed by the reaction process and an overlayer of VO$_x$ arising from diffusion of oxygen from the V$_3$Si/SiO$_2$ reaction boundary toward the surface.
 
\section{Results and discussions}
\subsection{Rutherford backscattering spectrometry analysis (RBS)}
RBS using a MeV \textsuperscript{4}He ions was employed to study the identity of the elements present in the sample and their composition as a function of depth. Here, we provide a brief overview of the technique's principles related to our study. A full description of the technique is provided in Ref.\cite{bib18}.
The energy of ions backscattered from elements at the sample surface is determined by their atomic mass. Ions backscattered from elements at the surface of the sample possess higher energy than those backscattered from elements deeper within the sample. This is due to the electronic stopping within the sample. Consequently, the thickness of a layer in the sample can be related with the energy width of the backscattered ions from that layer. Hence, the energy spectrum gives information about both the mass and the depth of the elements. 

The concentration or yield in RBS is proportional to the square of the atomic number (\(Z^2\)), making the technique less sensitive to light elements. This sensitivity decreases further when light element signals overlap with background signals from heavier elements. In our study, this limitation is particularly evident when detecting impurities such as carbon (C) and oxygen (O). For instance, in the phase-transformed VO$_x$ films shown in Figure \textcolor{blue}{2b} and \textcolor{blue}{2c}, the O signal appears broad and suppressed by the strong V background, which further diminishes the detectability of O.

\subsection{RBS results}
We conducted RBS analysis on several samples during and after the synthesis of V$_3$Si thin films, that is on the as-deposited V film (V/SiO$_2$/Si) and samples annealed at 750 °C and 800 °C for 30 minutes. RUMP RBS analysis \cite{bib19} was used to simulate and fit the experimental RBS data, for quantitative analysis of the elemental composition and thickness, as well as spectral interpretation.

Figure \textcolor{blue}{2a} displays the RBS spectra of the as-deposited V sample, showing signals from all relevant elements. The V spectrum is observed in the channel range of 320–400, and the Si signal is represented by two edges: one at the higher energy level (marked as \textbf{1}), which corresponds to the top Si face on the SiO$_2$ layer, and the other at the lower energy level (marked as \textbf{2}), which corresponds to the Si from the Si sustrate/ SiO$_2$ interface. The signal, marked as ``O/SiO$_2$'', corresponds to the O spectrum from the SiO$_2$ layer. The simulated (SIM) spectrum (shown in red) fits well the experimental (EXP) spectrum (in black), with the SIM data being presented in the table below Figure \textcolor{blue}{2a}. The thicknesses for the V and SiO$_2$ layers align well with the expected deposited V and the thermally grown SiO$_2$ thickness layer.
Figure \textcolor{blue}{2b} and \textcolor{blue}{c} show the RBS spectra of samples annealed at 750°C and 800°C for 30 minutes, respectively. In both figures, the simulated spectra (red) closely match the experimental spectra (black), apart from a deviation in the near-surface region of the VO$_x$ in Figure \textcolor{blue}{2c} which we have not tried to fit because we are not concerned with the exact composition of the VO$_x$ film with depth.
In the 750°C annealed sample (Figure \textcolor{blue}{2b}), a V$_3$Si layer approximately 63 nm thick forms beneath a 300 nm thick VO$_x$ layer. The SiO$_2$ layer also decreases in thickness from 230 nm to 170 nm, indicating the consumption of Si and O for the formation of V$_3$Si and VO$_x$. These changes are more pronounced in the 800°C annealed sample (Figure \textcolor{blue}{2c}). The higher annealing temperature enhances V diffusion and accelerates SiO$_2$ decomposition, leading to thicker V$_{3}$Si and VO$_x$ layers. Quantitative RBS analysis of this 800°C annealed sample reveals that the V$_{3}$Si layer grows to about 130 nm, the VO$_x$ layer reaches  $\sim$ 315 nm, and the residual SiO$_2$ layer is reduced to $\sim$ 75 nm. Overlay of the EXP spectra of the as-deposited (black) and 750°C-annealed (red) samples in Figure \textcolor{blue}{2d}  highlights three key observations:
\noindent \textbf{I.} A decrease in the yield of the vanadium (V) signal (marked as \textbf{1a}) and the broadening of its spectrum (marked as \textbf{1b}) associated with V diffusion into the Si/SiO$_{2}$ substrate, forming V$_{3}$Si at the interface of the SiO$_{2}$ layer. 
\noindent \textbf{II.} The reduced yield of the Si signal (marked as \textbf{2a}) and the step formation towards the V spectrum (marked as \textbf{2b}) indicate Si reaction with V, forming V$_{3}$Si. 
\noindent \textbf{III.} The O profile of the SiO$_{2}$ layer shows a reduction in yield at its higher energy levels (marked as \textbf{3a}), and a broader peak (marked as \textbf{3b}) emerging next to the Si signal is associated with the O in VO${x}$. The sharp decline in the V signal yield (marked as \textbf{1a}) further supports the formation of VO$_{x}$, consistent with V readily forming oxide compounds.\cite{bib8}

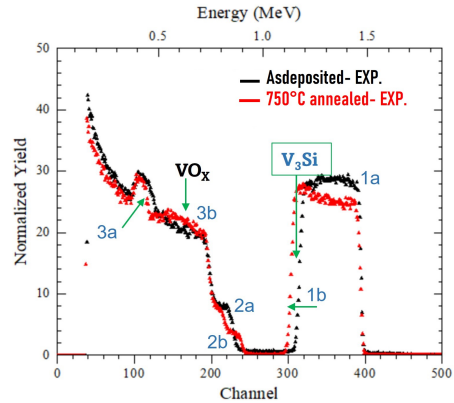
\begin{figure*}[h!]  
    \centering
    \begin{tikzpicture}
        % First row subfigures (a) and (b)
        \node[anchor=north west] (a) at (0,-0.4) {
            \begin{subfigure}[b]{0.43\textwidth}
                \centering
                \includegraphics[width=\textwidth]{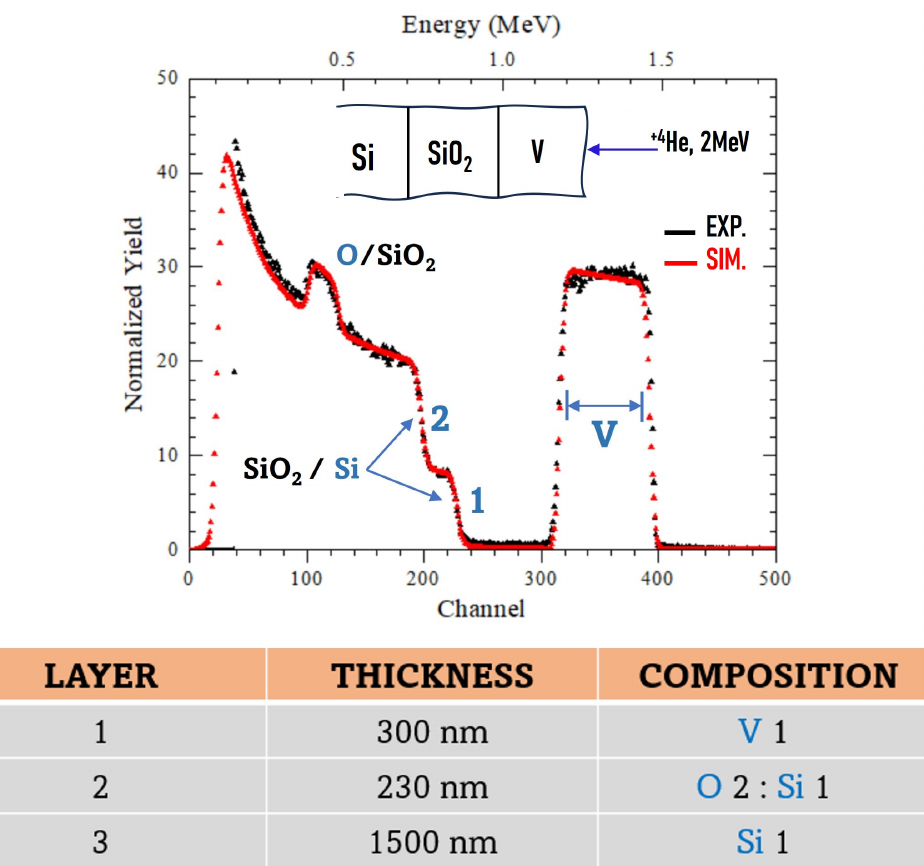}
                \label{fig:2a}
            \end{subfigure}
        };
        \node[anchor=north west] (b) at (8.7,-0.4) {
            \begin{subfigure}[b]{0.43\textwidth}
                \centering
                \includegraphics[width=\textwidth]{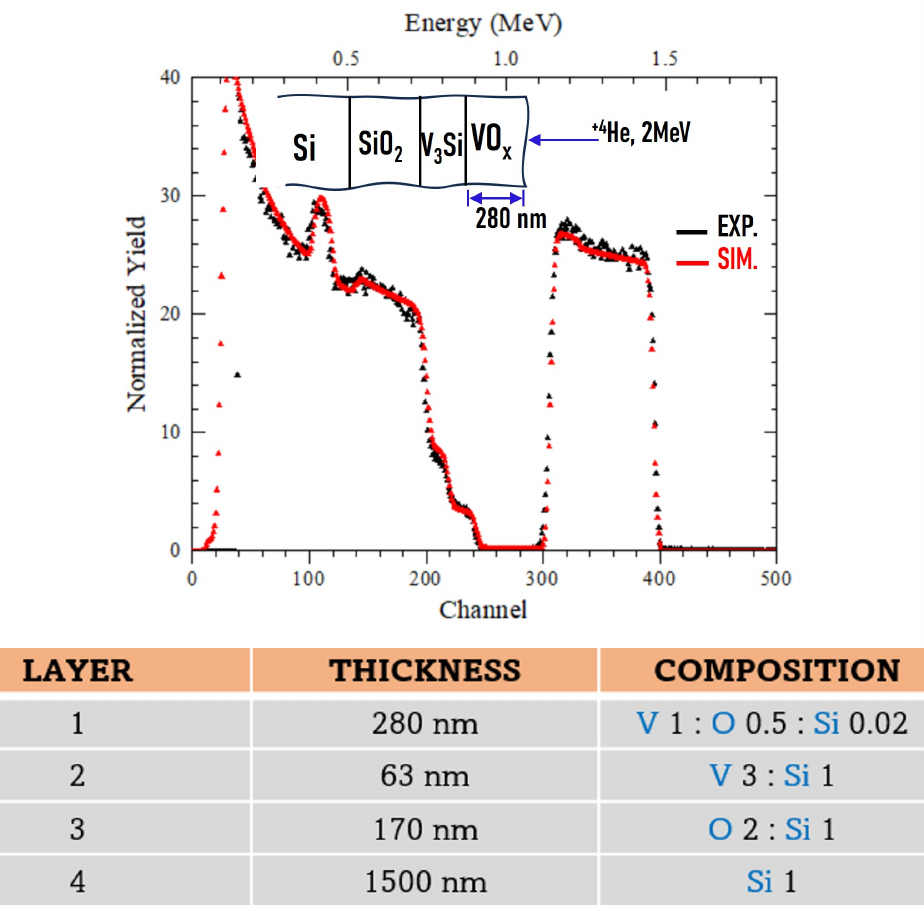}
                \label{fig:2b}
            \end{subfigure}
        };

        % Labels (a), (b)
        \node[anchor=north west, yshift=1.6em, xshift=1em] at (a.north west) {\large \textbf{(a) Asdeposited-V film}};
        \node[anchor=north west, yshift=1.6em, xshift=1em] at (b.north west) {\large \textbf{(b) Annealed @ 750°C}};

        % Second row subfigures (c) and (d)
        \node[anchor=north west] (c) at (0.3,-9.8) {
            \begin{subfigure}[d]{0.44\textwidth}
                \centering
                \includegraphics[width=\textwidth]{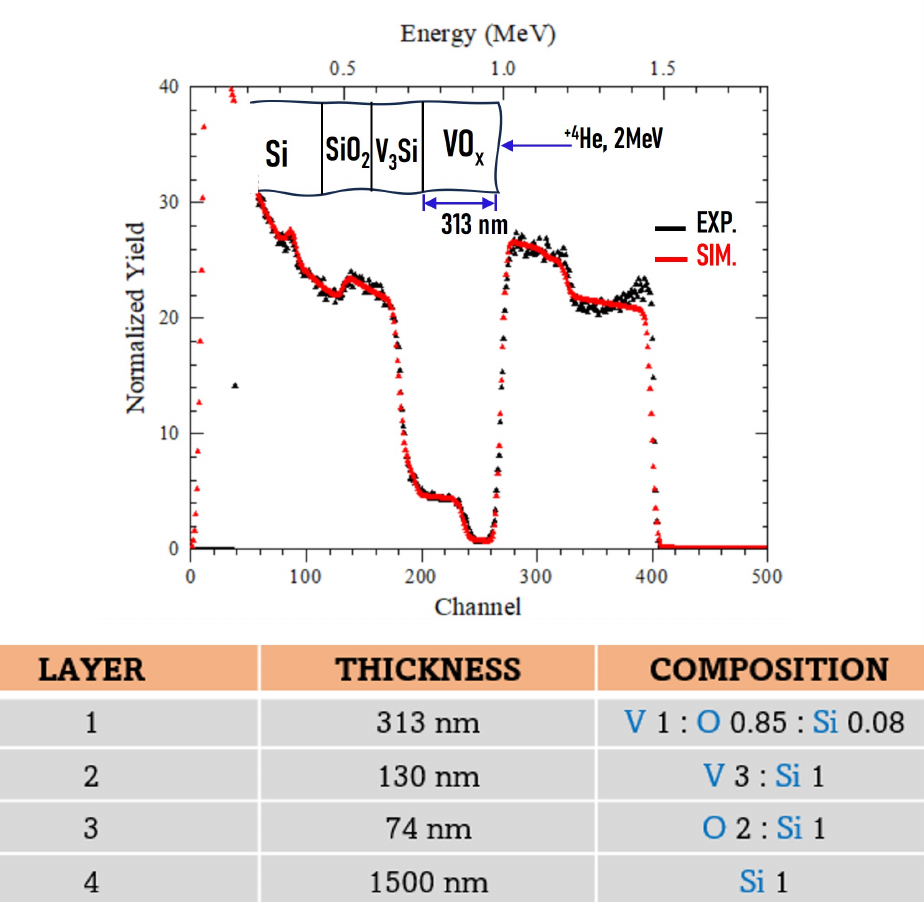}
                \label{fig:2c}
            \end{subfigure}
        };
        \node[anchor=north west] (d) at (9.3,-9.8) {
            \begin{subfigure}[d]{0.325\textwidth}
                \centering
                \includegraphics[width=\textwidth]{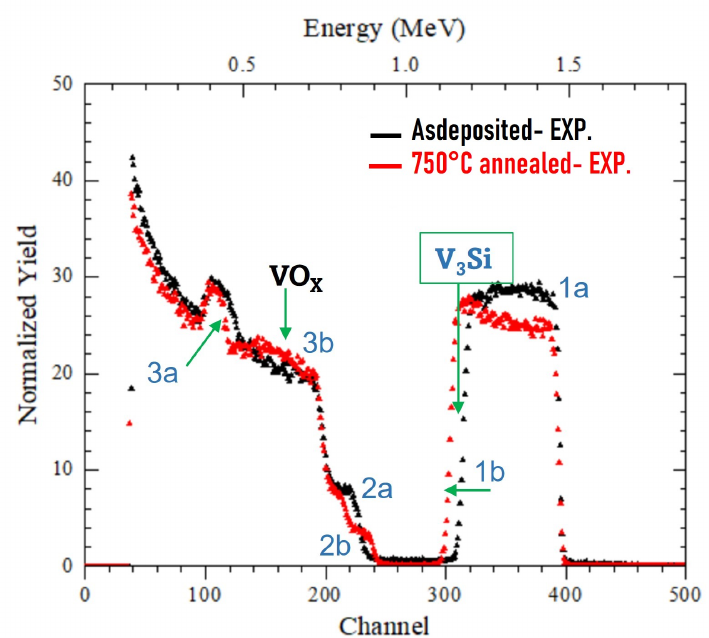}
                \label{fig:2d}
            \end{subfigure}
        };

        % Labels for second-row (c), (d)
        \node[anchor=north west, yshift=1.4em, xshift=1em] at (c.north west) {\large \textbf{(c) Annealed @ 800°C}};
        \node[anchor=north west, yshift=1.4em, xshift=-1.2em] at (d.north west) {\large \textbf{(d) Overlay Spectra}};

     \end{tikzpicture}
    
    \caption{RBS spectra of the V/SiO$_2$/Si samples analyzed using RUMP software, showing signals of the possible elements in the sample as a function of depth: (\textbf{a}) Spectra of the as-deposited V/SiO$_2$/Si film. (\textbf{b}) and (\textbf{c}) Spectra of the samples annealed at 750°C and 800°C for 30 minutes, respectively, showing different signals, including the formation of V$_3$Si at the SiO$_2$-VO$_x$ interface. For quantitative analysis, a table of simulation data that corresponds to the experimental data plot is provided below the respective subfigures. (\textbf{d}) Overlay of the RBS spectrum for the as-deposited and 750°C annealed samples, highlighting the phase transformation to V$_3$Si, the VO$_x$ signal, and the overall changes in the V and SiO$_2$ phases following annealing.}
    \label{fig:Figure_2}
\end{figure*}

% Introduce text placement flexibility right after figure:

\subsection{Low temperature electrical resistance measurement}
Resistance measurements using the four-probe method in a van der Pauw (VDP) configuration were conducted on the V$_3$Si thin film annealed at 750°C to examine its temperature-dependent electrical behaviour. The results, shown in Figure \textcolor{blue}{3a}, demonstrate that the sample exhibits a sharp SC transition, with a $T_c$ of 13~K. The steep resistance drop near the SC transition results in a narrow transition width of $\Delta T_c = 0.6$~K. This suggests a high degree of homogeneity of the V$_3$Si thin film. and a good interconnection between the polycrystalline grains of the SC thin film in agreement with previous studies.\cite{bib31, bib32}

Having successfully formed V$_3$Si thin films on thermally grown SiO$_2$/Si substrates and demonstrated their SC properties,we proceeded to fabricate SC patterned wires for applications such as quantum computing. Using standard photolithography techniques, we patterned wires with widths ranging from 5~$\mu$m to 200~$\mu$m, processing them under conditions similar to those used for the thin films. However, electrical contacts to the SC layer that functioned at low temperature could not be reliably formed in the patterned devices and it was considered that this may be due to the presence of the VO$_x$ overlayer.
\subsection{Effect of \texorpdfstring {VO$_x$}: on the Superconducting Layer}

It is not entirely clear whether the vanadium oxide (VO$_x$) formed at the surface layer is responsible for insulating the electrical contact to the V$_3$Si in the case of patterned wires. However, studies have shown that annealing V/SiO$_2$/Si films promotes the formation of not only vanadium silicides but also various VO$_x$ phases, including VO, V$_2$O$_3$, VO$_2$, and V$_2$O$_5$,\cite{bib7, bib8, bib9, bib13, bib14} each exhibiting distinct electrical properties. For example, VO$_2$ undergoes a metal-insulator transition (MIT) at approximately 341 K, transitioning from a metallic to an insulating state as the temperature decreases.\cite{bib23} Interestingly, oxygen-deficient VO$_{2-x}$ films have been observed to maintain metallic conductivity even at temperatures as low as 1.8 K, effectively suppressing the MIT.\cite{bib20, bib24} 
Furthermore, V\(_4\)O\(_7\)\cite{bib26} and V\(_6\)O\(_{11}\)\cite{bib27}, known as the Magnéli phases (denoted as V\(_n\)O\(_{2n-1}\)) of VO\(_x\) compounds\cite{bib25}, exhibit intriguing electrical properties due to their mixed metallic and semiconducting nature. Specifically, these compounds can display metallic conductivity even at cryogenic temperatures.
In contrast, phases with higher oxygen content, such as V$_2$O$_5$, are known to be insulating at low temperature.\cite{bib21, bib22} Even the VO$_2$ exhibits insulating behavior around the SC transition temperature of V$_3$Si.

While the larger contact pads used for the VDP structures on the laterally uniform thin films did reliably allow low temperature electrical characterisation of the SC transition, the smaller scale of the pads on the patterned devices resulted in variability in behaviour from contact to contact and it was surmised that local grain structure and compositional variation of the VO$_x$ layer could be responsible for this issue.

Therefore, we decided to investigate the removal of the VO$_x$ layer to expose the V$_3$Si phase. In this study, we used RIE to etch away the VO$_x$ layer. The preliminary results of this process are discussed in the next section.

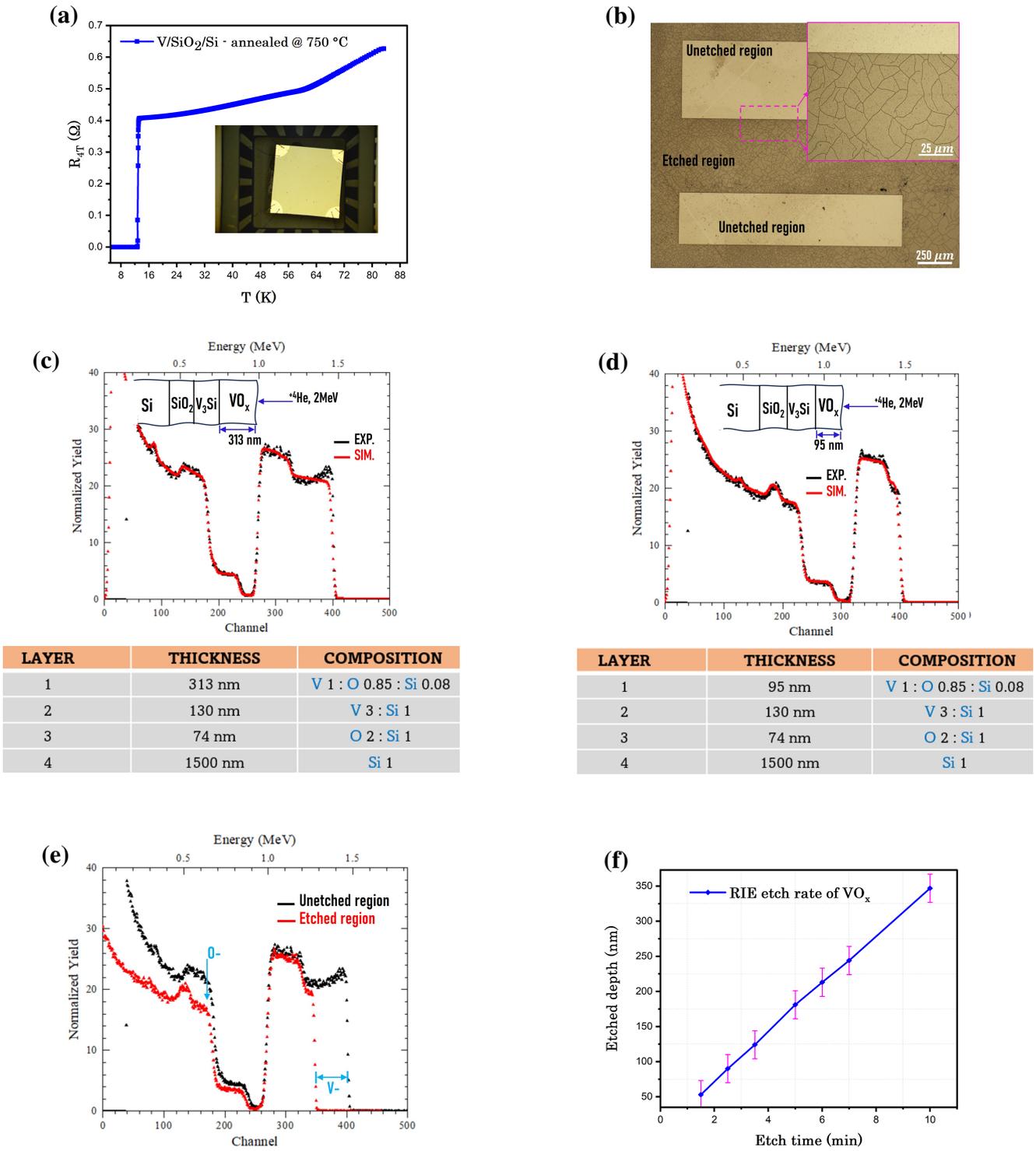
\begin{figure*}[h!]  
    \centering
    \begin{tikzpicture}
        % First row subfigures (a) and (b)
        \node[anchor=north west] (a) at (0.3,0.25) {
            \begin{subfigure}[b]{0.4\textwidth}
                \centering
                \includegraphics[width=\textwidth]{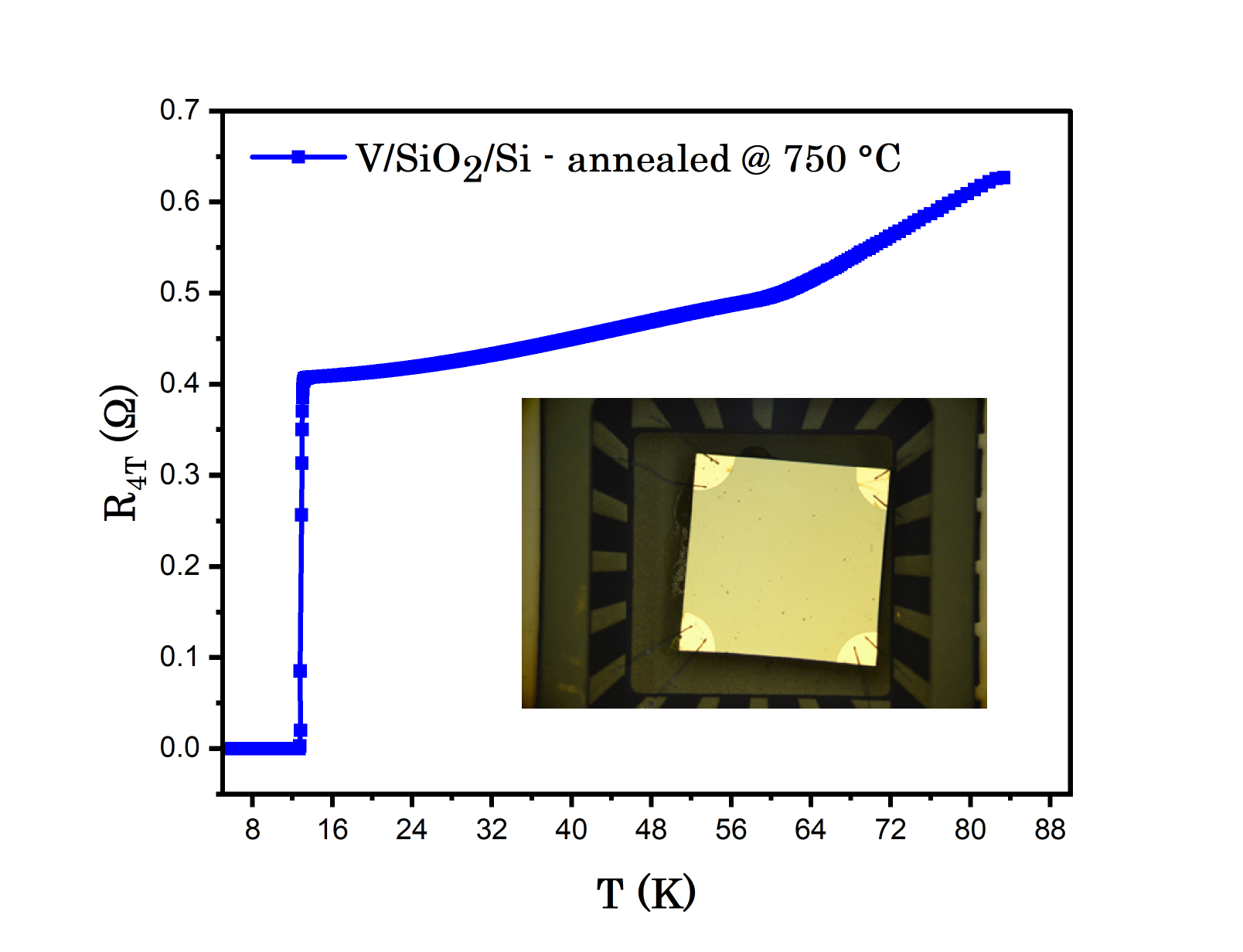}
                \label{fig:3a}
            \end{subfigure}
        };
        \node[anchor=north west] (b) at (10.75,-0.35) {
            \begin{subfigure}[b]{0.29\textwidth}
                \centering
                \includegraphics[width=\textwidth]{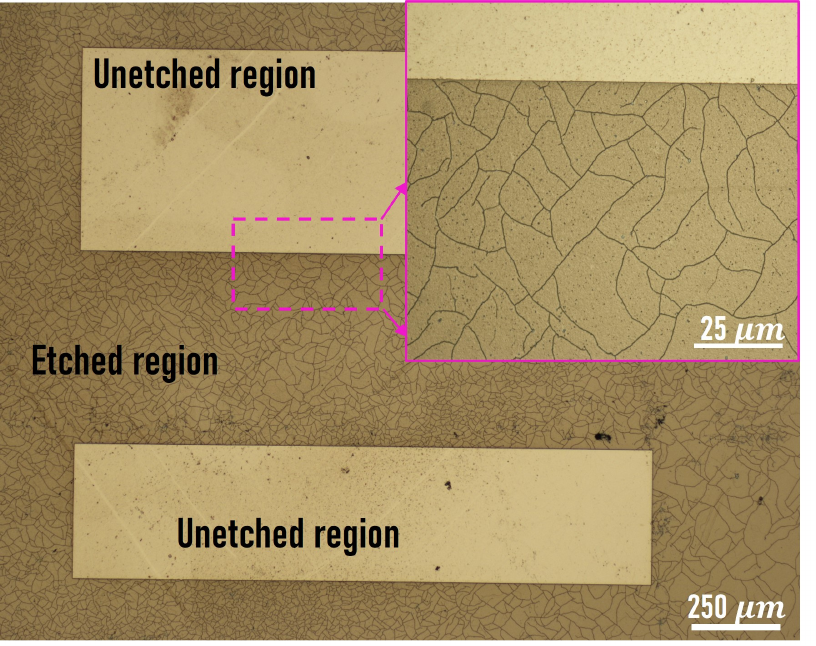}
                \label{fig:3b}
            \end{subfigure}
        };

        % Labels (a), (b)
        \node[anchor=north west, yshift=-0.9em, xshift=0.8em] at (a.north west) {\large \textbf{(a)}};
        \node[anchor=north west, yshift=0.8em, xshift=-2.2em] at (b.north west) {\large \textbf{(b)}};

        % Second row subfigures (c) and (d)
        \node[anchor=north west] (c) at (-0.2,-5.6) {
            \begin{subfigure}[b]{0.42\textwidth}
                \centering
                \includegraphics[width=\textwidth]{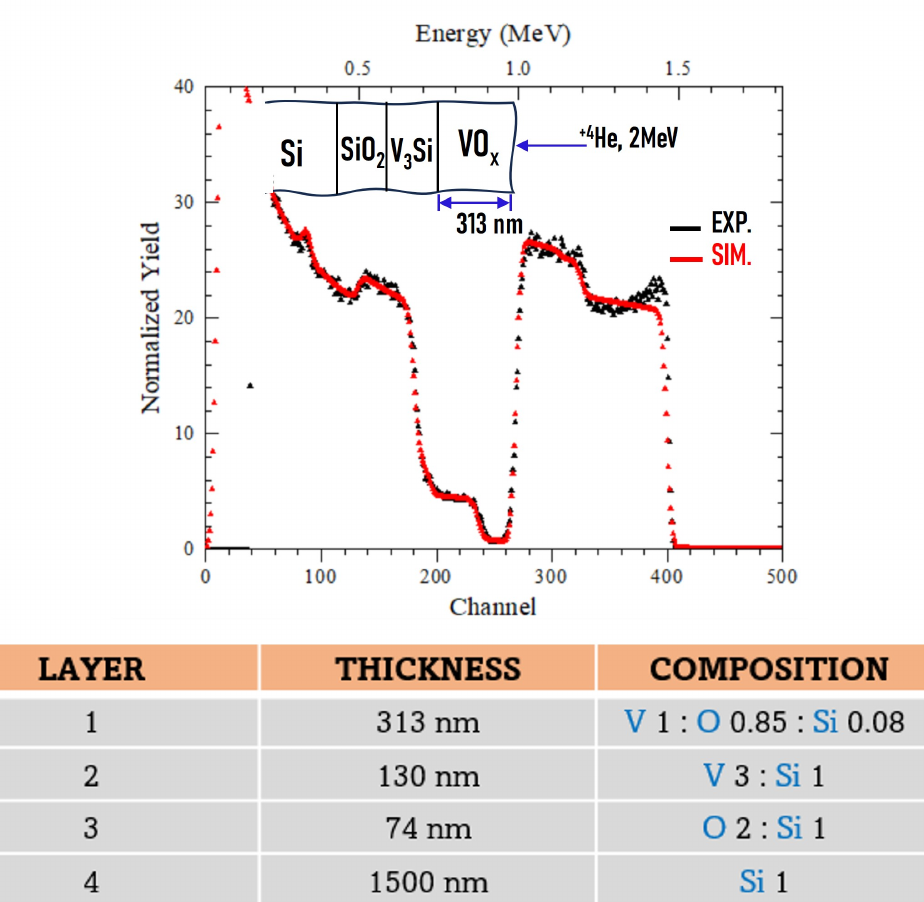}
                \label{fig:3c}
            \end{subfigure}
        };
        \node[anchor=north west] (d) at (9.5,-5.6) {
            \begin{subfigure}[b]{0.42\textwidth}
                \centering
                \includegraphics[width=\textwidth]{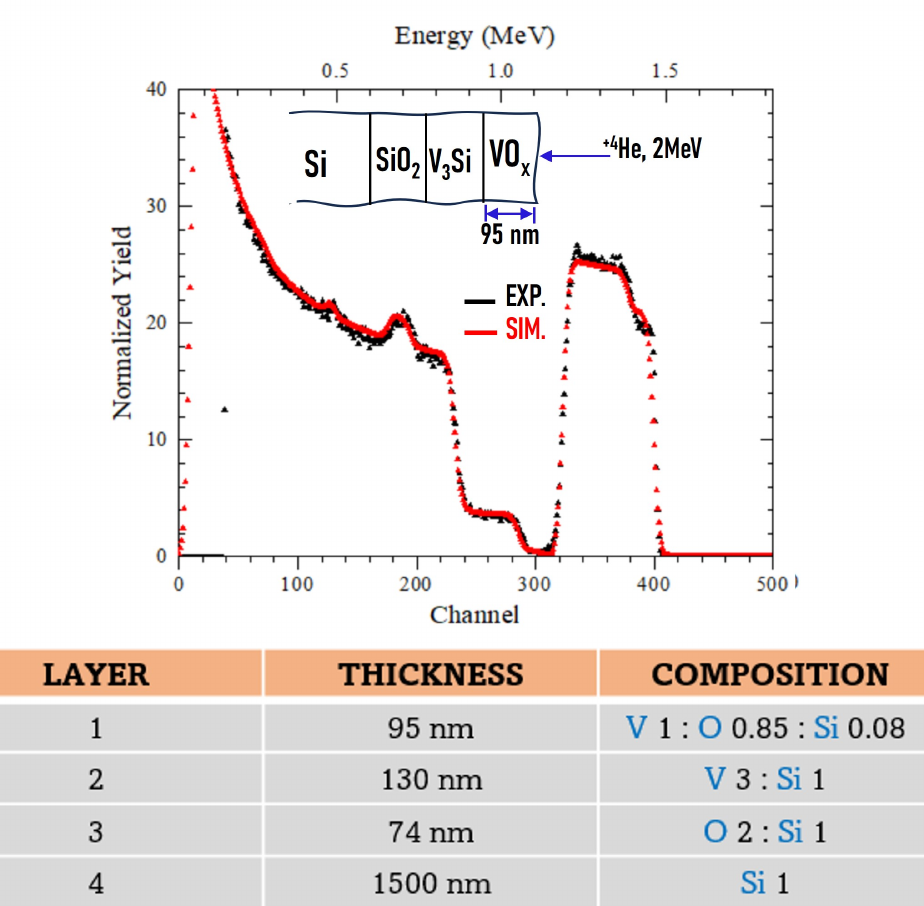}
                \label{fig:3d}
            \end{subfigure}
        };

        % Labels for second-row (c), (d)
        \node[anchor=north west, yshift=-1em, xshift=1.4em] at (c.north west) {\large \textbf{(c)}};
        \node[anchor=north west, yshift= -1.05em, xshift=1em] at (d.north west) {\large \textbf{(d)}};

     % third row subfigures (e), (f)
        \node[anchor=north west] (e) at (0.8,-14.05) {
            \begin{subfigure}[e]{0.33\textwidth}
                \centering
                \includegraphics[width=\textwidth]{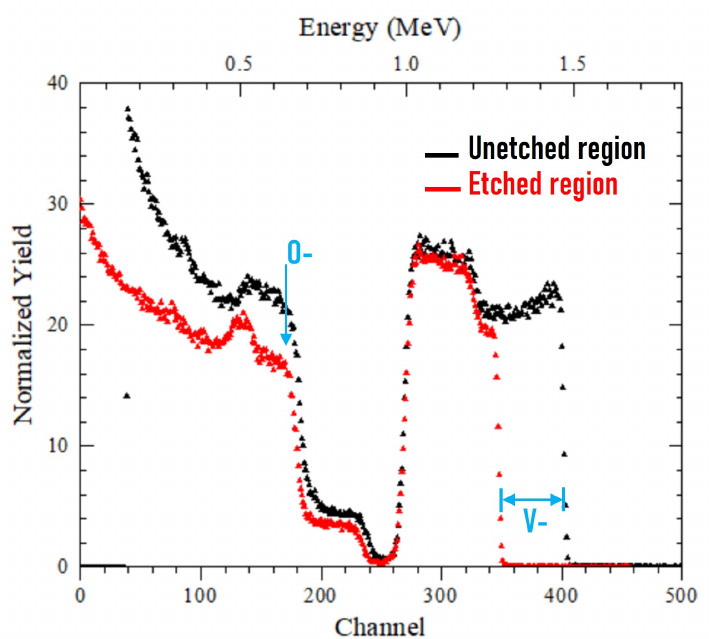}
                \label{fig:3e}
            \end{subfigure}
        };
            \node[anchor=north west] (f) at (9.6,-14.14) {
            \begin{subfigure}[f]{0.4\textwidth}
                \centering
                \includegraphics[width=\textwidth]{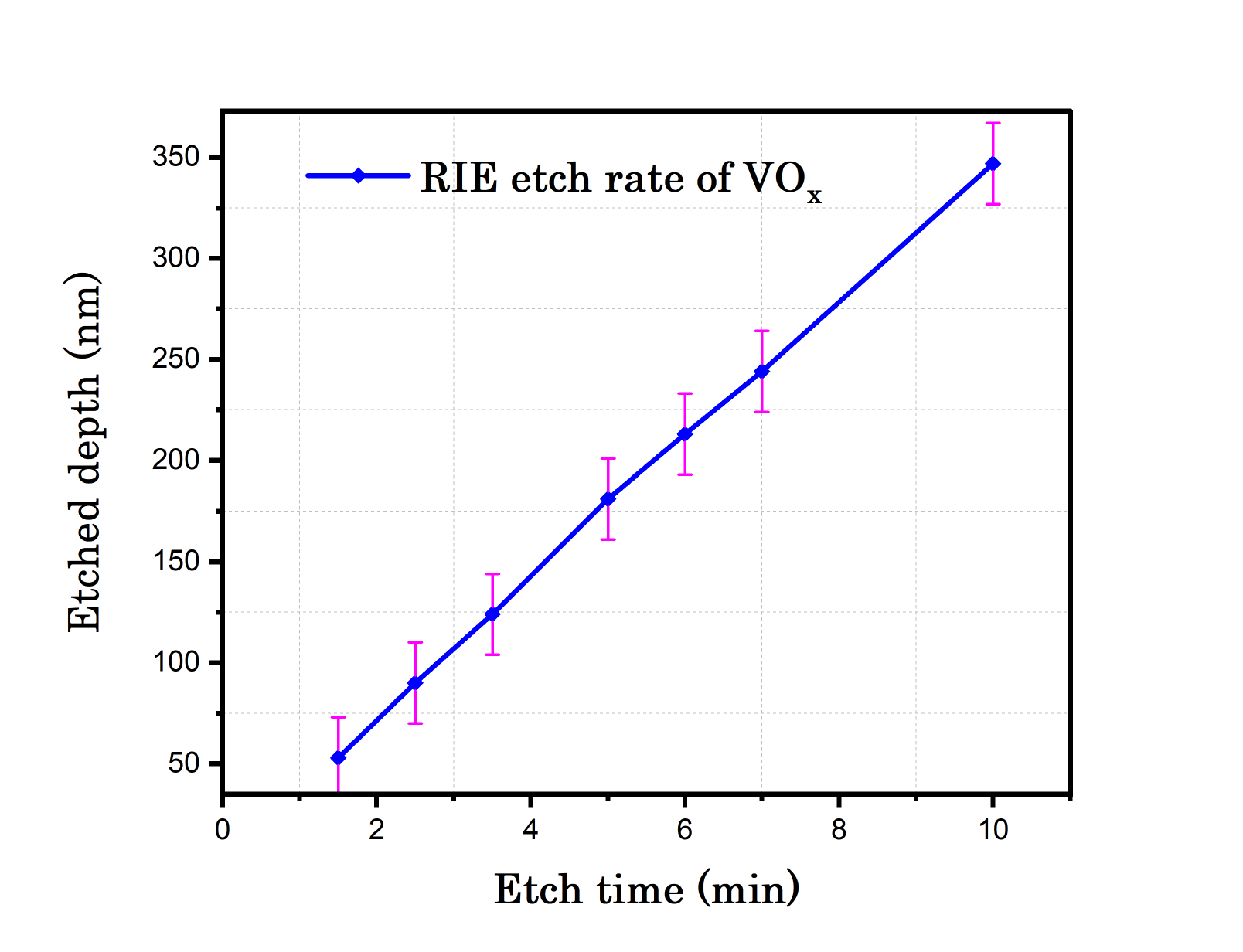}
                \label{fig:3f}
            \end{subfigure}
        };
        % Labels for third-row (e)
        \node[anchor=north west, yshift=-0.9em, xshift=-1em] at (e.north west) {\large \textbf{(e)}};
         \node[anchor=north west, yshift=-0.9em, xshift=1em] at (f.north west) {\large \textbf{(f)}};
        
    \end{tikzpicture}
   \vspace{-3pt}
\caption{\textbf{(a)} Four-terminal resistance as a function of temperature for a thin film sample annealed at 800°C, with its micrograph shown in the inset.(\textbf{b}) Optical micrograph of the sample highlighting the surface difference between the etched and unetched regions. \textbf{(c-e)} RBS data analysis of the RIE-etched and unetched regions of the 800°C annealed thin film sample: \textbf{(c)} RBS spectra from the unetched region of the sample with table of data collected from its SIM plot. \textbf{(d)} RBS spectra from the etched region of the sample. The quantitative analysis from the SIM (in the table below) indicates that about 208 nm of the surface layer (VO\(_x\)) was removed after 6 minutes of RIE with 95 nm remaining. \textbf{(e)} Overlay of the experimental RBS data from the etched and unetched regions, showing the reduced VO\(_x\) spectra on the surface. \textbf{(f)} Optical profilometer analysis of the RIE etching rate from seven tests, showing a consistent etching rate of \(34.51 \pm 0.39\) nm/min.}
    \label{fig:Figure_3}
\end{figure*}

\section{Reactive ion etching of VO$_x$ film layer}
RIE is a technique that utilizes chemically reactive plasma to etch specific materials. Research has shown that adjusting parameters such as radio frequency (RF) power, gas composition, and pressure, the etch rate and selectivity for VO$_x$ films can effectively control the etch rate and selectivity for VO$_x$ films, achieving rates of over 74 nm/min.\cite{bib28,bib29} However, unlike these VO$_x$ films which are grown by deposition, our VO$_x$ films are phase-transformed and sitting on a SC V$_3$Si layer. With the goal of selectively etching VO$_x$ while leaving the V$_3$Si layer intact, a careful and gradual etching process is required to preserve the integrity of the underlying V$_3$Si layer and RBS provides a convenient tool for studying the etch process and to identify any compositional changes that may accompany the process.

In our RIE experiment, a mixture of Ar/Cl$_2$ gases, was employed to remove the VO$_x$ overlayer. In this process, a 4 $\times$ 4 mm\textsuperscript{2}, 800 °C-annealed thin film sample, consisting of a VO$_x$/V$_3$Si/SiO$_2$/Si multilayer structure, were patterned into two rectangular shapes with dimensions of 1.7 mm $\times$ 0.6 mm and 1.7 mm $\times$ 0.4 mm. This patterning was achieved using direct laser writing (DLW) after the sample was coated with approximately 3 $\mu$m of Ti-35E photoresist (PR).
The operating parameters included a total gas pressure of 5 mTorr, an input ICP power of 1500 W, a bias power (RIE power) of 200 W, and a total gas flow rate of 40 sccm, with varying etching durations. The Cl$_2$/Ar mixing ratio was maintained at 25\% Cl$_2$ and 75\% Ar by adjusting the partial flow rates of each component. The bottom electrode temperature was stabilized at 14 °C using a water-cooled system. Etch depths were measured using a stylus profilometer (Bruker Dektak XT). 

An optical micrograph of the etched and unetched regions of the 800°C-annealed V/SiO$_2$/Si sample in Figure \textcolor{blue}{3b} displays a smooth surface, indicating a uniformly etched surface and the granular structure in the etched region is associated with the residual VO$_x$ as confirmed by the RBS analysis in Figure \textcolor{blue}{3d}.
RBS analysis was performed on both the etched and unetched regions of the sample (see Figure \textcolor{blue}{3c-e}), and profilometer measurements were used to further analyse the etched depth (Figure \textcolor{blue}{f}). 

Figure \textcolor{blue}{3c} displays the RBS analysis of the unetched region of the 800 °C-annealed sample, which corresponds to the data presented in Figure \textcolor{blue}{2c}. The results presented in the table show that a 130 nm V$_3$Si phase is formed underneath $\sim$ 315 nm VO$_x$ layer. Figure \textcolor{blue}{3d} illustrates the RBS spectra analysis of the etched region, with the SIM plot (red) closely matching the experimental data (black). The analysis shows that approximately 95 nm thick of VO$_x$ remains at the surface layer, and this reflects the oxide layer being reduced by about 208 nm after the 6-minute RIE etching. 
Overlaying the RBS spectra of the etched (red plot) and unetched (black plot) regions of the 800 °C-annealed sample shown in Figure \textcolor{blue}{3e}, demonstrates the reduction of the VO$_x$ layer from both the V and O spectral regions. 
Profilometer measurements of the 800 °C-annealed thin film sample showed that $208 \pm 5$ nm of depth was etched after 6 minutes, closely matching the RBS analysis. A slower etching rate was targeted to avoid rapid etching into the V$_3$Si layer. Figure \textcolor{blue}{3f} presents the RIE etching rate of VO$_x$ films from similarly processed samples, showing a consistent rate of \(34.51 \pm 0.39\) nm/min. This demonstrates good process control, which is crucial for the next phase of the project—fabricating V$_3$Si devices from these films.

These initial test results are encouraging and support the continuation of this approach to fully etch the VO$_x$ while maintaining the integrity of the underlying V$_3$Si layer.

\section{Conclusions and outlooks}
V$_3$Si SC thin films were fabricated by electron beam evaporation of V onto a thermally grown SiO$_2$/Si substrate, followed by HVEBA. A sharp resistivity drop and a complete SC transition were observed with a T$_c$ of 13 K for the sample annealed at 750°C. The thickness and composition of the multilayer thin film were studied using RBS analysis.

RIE was used to remove the phase-transformed VO$_x$ surface layer, exposing the V$_3$Si layer as a step toward the creation of functional patterned SC devices. 

With further tuning of the RIE parameters we expect that it will be possible to reliably making electrical contacts to the SC thin film layer allowing reliable construction of patterned devices and providing a pathway for development of functional SC V$_3$Si thin film devices including Josephson junctions and SC resonators.

\section*{Acknowledgements}
This work was supported by an Australian Research Council (ARC) grant (DP200103233). It was also conducted, in part, at the Melbourne Centre for Nanofabrication (MCN) in the Victorian Node of the Australian National Fabrication Facility (ANFF). We acknowledge the AFAiiR node of the NCRIS Heavy Ion Capability for providing access to ion-beam analysis facilities. Fshatsion Gessesew is grateful for the Melbourne International Fee Remission Scholarship (MISRF) and the Melbourne International Research Scholarship (MIRS). The authors would like to thank S. Gregory and R. Szymanski for their technical assistance in maintaining the Pellerton and clean room instruments, ensuring their safe operation.

\bibliographystyle{elsarticle-num} 
\bibliography{References}

\end{document}